# High On/Off Ratio Graphene Nanoconstriction Field Effect Transistor

Ye Lu[†], Brett Goldsmith[†], Douglas R. Strachan[†‡], Jonghsien Lim[§], Zhengtang Luo[†], A.T. Charlie Johnson[*†]

[†] Department of Physics and Astronomy, University of Pennsylvania, Philadelphia, PA, 19104

[‡]Department of Physics and Astronomy, University of Kentucky, Lexington, KY, 40506

[§]Department of Physics, Swarthmore College, Swarthmore, PA 19081

*Corresponding author, cjohnson@physics.upenn.edu

**We report a method to pattern monolayer graphene nanoconstriction field effect transistors (NCFETs) with critical dimensions below 10 nm. NCFET fabrication is enabled by the use of feedback controlled electromigration (FCE) to form a constriction in a gold etch mask that is first patterned using conventional lithographic techniques. The use of FCE allows the etch mask to be patterned on size scales below the limit of conventional nanolithography. We observe the opening of a confinement-induced energy gap as the NCFET width is reduced, as evidenced by a sharp increase in the NCFET on/off ratio. The on/off ratios we obtain with this procedure can be larger than 1000 at room temperature for the narrowest devices; this is the first report of such large room temperature on/off ratios for patterned graphene FETs.**

Keywords: graphene, transistor, nanoconstriction, fabrication

*Introduction*

Graphene, a single layer of carbon atoms organized in a honeycomb crystal lattice, has drawn tremendous attention for its extraordinary electronic properties[1], among other attributes. It is reported to have higher electron mobility than any known semiconductor,[2-4] which makes it an excellent candidate for next generation analog and digital electronics.

Moreover, graphene is distinguished among other nanomaterials by its potential for use in "top-down" fabrication of integrated circuits[5]. Transistor applications in digital electronics require the induction of an energy gap in graphene, which in the bulk form (i.e., with lateral dimensions exceeding 1 μm) is naturally a two-dimensional, zero band gap semimetal. An energy gap can be created through quantum confinement and/or edge effects by patterning graphene in the form of a nanoribbon[6, 7], quantum dot[8] or, as shown here, a nanoconstriction. The exact shape and structure of the graphene edge is expected to lead to different band gaps for similarly sized structures[9-15], and the effective electronic gap should be greatly influenced by the presence of disorder[9, 12, 16-18]. This effective gap is reflected in the on/off ratio measured in a field effect transistor structure; to create a useful device this ratio should be large, minimizing off state current and maximizing potential gain.

Multiple approaches towards the fabrication of nanoscale graphene structures have been reported, including conventional nanolithographic patterning[19, 20], opening of carbon nanotubes to create narrow graphene sheets[21, 22], crystallographic etching of graphene to form nanoribbons using metal particles[23-27] or a scanning tunneling microscope(STM)[28], and chemical routes to isolate small structures from bulk graphite[6]. State-of-the-art lithographic techniques are limited to a resolution of 10-20nm, and thus it is challenging to make "top-down" graphene devices below this scale[20]. To date, the largest room-temperature on/off ratios reported for patterned graphene devices are less than 100 [19, 20]. Crystallographic graphene nanoribbons formed by etching with metal particles or STM are of great scientific interest; however, the control of such ribbons is challenging, and measurements of FET devices based on these approaches are yet to be reported. "Bottom-up" approaches have also been used to create graphene nanoribbon devices with large on/off ratios[6, 19, 21, 22]. Electrical measurements on these devices have shown that sub-10nm graphene nanoribbons function as hole-transport FETs with on-off ratios approaching $10^7$. Despite this progress, it remains to be shown that FET devices with suitably large on/off ratios ($I_{on}/I_{off}$) can be directly fabricated using lithographic approaches or some other "top-down" procedure.

When metal contacts are deposited on graphene, Schottky barriers form at this interface. Small area contacts to graphene nanoribbons are high resistance and readily depleted by the backgate; thus the electrical resistance of graphene nanoribbons with such contacts is dominated by the Schottky barriers rather than the graphene channel itself.[7] In contrast, the contact between a metal and large-area "bulk" graphene can be nearly ohmic with resistance below 1 kΩ[19]. It is therefore desirable to investigate device geometries with narrow channels and large area contacts, so the resistance is dominated by conduction through the graphene channel rather than the contact region. Notably, electron beam lithography has been used to create large width graphene nanoribbons with nearly ohmic contacts[19, 20].

To create small-width graphene devices with low contact resistance, we used a combination of feedback controlled electromigration (FCE)[29, 30] and e-beam lithography to controllably fabricate nanoconstriction field effect transistors (NCFETs) on scales below the traditional limits of nanolithography. Interestingly, theoretical models of the effect of edge localization suggest that creating constriction based FETs should lead to a larger effective energy gap than similarly sized ribbons[14]. We fabricated NCFETs with constriction widths in the range 8 – 150 nm, as shown in Figure 1, with the narrowest devices exhibiting properties similar to those observed for sub-10nm graphene nanoribbons, including on/off ratios higher than $10^3$ at room temperature. Moreover, the metal contacts to the NCFETs are much larger than what is possible for chemically derived nanoribbon devices[6], and are thus "bulk" contacts, whose resistance is much smaller than that of the graphene channel. This allows a direct measurement of the electrical characteristics of the graphene nanoconstriction with only a minor contribution from any contact barriers.

*Results*

We find that it is possible to perform the FCE process on a e-beam patterned gold structure on top of conductive graphene. During FCE, a slowly ramped voltage is used to electromigrate gold out of a constriction in a "bow tie" structure. Figure 1(b) shows a typical FCE voltage curve for one of our devices. As the gold is thinned, a lower voltage is needed to continue electromigration. A computer program monitors the change in conductance and lowers the electromigration voltage in steps as the gold structure is thinned. In the typical literature on FCE, this process is continued until a tunneling gap is opened[29, 30]. We stop the FCE process here while a small ~10nm junction of gold is left in the middle of the structure. Figures 1(c) and 1(d) are scanning electron microscope (SEM) images of a sample after the FCE process showing the very narrow gold constriction on top of monolayer graphene. Following FCE, an oxygen plasma etch is used to remove graphene regions that are unprotected by the gold, and then the sample is immersed in a $KI/I_2$ gold etch solution for washing away the remaining gold structure.

Typical data from devices made using this FCE-based approach are shown in Figures 2 and 3. Figure 2(a) is an SEM micrograph of a device with constriction width of 90nm±5nm, which has a current on/off ratio ($I_{on}/I_{off}$) of 14 at room temperature (Figure 2(b)). Figure 2(c) is a semi-logarithmic plot of the current vs. gate voltage (I-Vg) characteristics of the same device at different bias voltages, demonstrating that the ambipolar behavior and current on/off ratio both are maintained for bias voltage in the range 0.5 – 5 mV. The current vs. bias voltage characteristics at various gate voltages are shown in Figure 2(d).

Figure 3(a) shows an I(Vg) curve for a device with an $I_{on}/I_{off}$ of about 17 in an ambient environment, while Figure 3(b) is the SEM image of the same device, showing that the constriction width is 37 nm±3nm. Similarly, the I(Vg) curve of a device with $I_{on}/I_{off}$ of 1100 is shown in Figure 3(c), while the corresponding electron micrograph (Figure 3(d)) shows a device width of 9 nm±3nm. Since it involves a risk of device contamination and damage, SEM imaging was conducted after all electronic measurements had been performed. Raman spectra obtained from these two devices after FCE are shown in Figure 4, and both demonstrate a high intensity, symmetric 2D band centered at ~ 2670 $cm^{-1}$, a feature that is commonly used to identify monolayer graphene.[31] The insets in Figure 4 show Lorentz fits to the 2D peaks, simply demonstrating the symmetric nature of the peaks. The small but significant D peak that is observed is indicative of atomic disorder, most likely due to the edges of the graphene NCFET as the Raman spot size (1μm) is much larger than the NCFET channel. In addition, the graphene is stable in vacuum during the FCE process (see experimental section) therefore the FCE process wouldn't change the property of graphene.

Devices with constriction widths below 10 nm deliver up to ~100 μA/μm at $V_b$=50mV, with an on-state resistance of about 50 kΩ (Fig. 3c). In contrast to measurements of chemically derived sub-10nm graphene nanoribbons where Schottky barrier effects dominate,[7] our devices have micrometer-scale source/drain contacts, so the measured resistance reflects the properties of the graphene nanoconstriction itself. Thus all NCFET devices measured exhibit a very small on state resistance (50kΩ at 9 nm device width) and ambipolar conduction similar to that of bulk graphene. Current fluctuations seen in Fig. 3(c), which are similar to those observed in chemically derived graphene nanoribbon FETs[6], could be due to quantum interference effects in this mesoscopic system,[8] or changes in electron scattering associated with position fluctuations of an unstable edge atom.

In examining monolayer graphene NCFETs with constriction widths ranging from sub-10nm to 100nm, we find that $I_{on}/I_{off}$ increases as the constriction width decreases (Figure 5). This trend indicates the opening of a transport energy gap and is consistent with other work on graphene nanoribbon FETs fabricated by both conventional lithography and chemical methods. Precise determination of the energy gap is difficult given the status of theoretical understanding (see next paragraph). A rough estimate of the gap may be obtained using the formula $I_{ON} / I_{OFF} \propto e^{E_g / k_B T}$, which best applies to the case of Schottky barrier dominated conduction. This would correspond to an energy gap as large as 180meV for our narrowest device, which is close to the usual value measured by other groups[6, 32].

Demonstrating the future potential of this technique, a single flake of graphene was fabricated into two NCFET devices by a recently developed parallel etching process[33]. In this case, a single FCE procedure was used, and the self-limiting nature of electromigration ensures that multiple parallel processes are performed at nearly the same rate. As the size of available high quality graphene continues to increase, this technique becomes more useful in generating arrays of patterned graphene transistors.

*Discussion*

There are a number of theoretical models that address the opening of an effective bandgap in graphene nanostructures. Models that focus on the role of disorder are more relevant to devices fabricated lithographically, including those discussed here, since there is little prospect that "top-down" fabrication methods used to date yield an atomically precise edge geometry. Modest levels of edge disorder are predicted to lead to the onset of Anderson localization in graphene nanoribbons.[20] A transport energy gap is induced even in otherwise metallic nanoribbons, resulting in little or no difference in the conduction of nanoribbons with different edge geometry.[9] Coulomb blockade of transport has recently been observed in graphene nanoribbon transport experiments at low temperature, [34,35, 36, 37] suggesting the formation of quantum dots along the length of the ribbon due to edge roughness,[38] random charged impurities near the graphene or structural deformation of the graphene itself.[39] An alternate view is that the existence of carrier density inhomogeneities ("electron-hole "puddles") associated with disorder from one of these sources, along with a confinement-induced energy gap in the nanoribbon, can lead to a percolation-driven metal-insulator transition.[39] Interestingly, calculations indicate that the conductance of graphene nanoconstrictions with "wedge" geometry similar to the NCEFT show pronounced suppression of transmission for electron states with energies less than that of the second carrier subband even for atomically perfect edges.[14] These authors trace this result to the fact that charge carriers are preferentially bound to the edges of the constriction and decoupled from the electrodes; this leads to an effective transport gap that is found to be robust against perturbations to the edge atomic structure. The calculated density of states for a constriction shows a large number of defect states, which may explain the current fluctuations we observe in our narrowest devices (see Figure 3(c)) [14, 40]. The movement of charge from regions capable of conduction to static edge states could also explain how such large $I_{on}/I_{off}$ ratios are achieved here[38, 39]. Future experiments to distinguish between these different models will be facilitated by progress towards structural control and/or structural characterization of the constriction with atomic resolution.

*Conclusions*

To summarize, using a gold etch mask narrowed by feedback controlled electromigration, we are able to fabricate monolayer graphene NCFETs with widths ranging from 100nm to sub-10nm. Devices with widths below 10 nm exhibit room temperature on/off current ratios larger than 1000, a level that is sufficient for use in advanced digital electronics. Additionally, sub-10nm devices have a relatively low on state resistance and show ambipolar behavior. These point contact like graphene NCFETs open up a route to verify theoretical predictions such as coherent electronic transport in subwavelength regime[41], and should provide a useful complement to measurements of graphene nanoribbons.

*Experimental Section*

Sample fabrication proceeds as follows. Mechanically exfoliated graphene sheets are deposited from kish graphite on an oxidized silicon substrate (300nm oxide thickness) with prefabricated Pd/Cr alignment markers. Monolayer graphene sheets are identified by optical microscopy and in some cases by atomic force microscopy (AFM) and Raman spectroscopy, as we previously reported[42]. Conventional electron beam lithography using PMMA C4 950 (Microchem) and thermal evaporation are used to define 20nm thick, 100nm wide bowtie shaped gold nanowires on top of monolayer graphene sheets. A second round of electron beam lithography and metal deposition is used to fabricate 60nm thick Pd/Cr source/drain leads that connect to the gold nanostructures on each side.

FCE is used to reproducibly narrow the gold bowtie structure by controlling the electromigration process that occurs under large current bias. The FCE process is performed on samples in a turbo-pumped vacuum chamber at a pressure below $5\times10^{-5}$ Torr, to ensure the stability of monolayer graphene at the elevated temperatures (200-350 C)[29, 43] generated by Joule heating during FCE[29]. During FCE we monitor the conductance of the sample while the voltage across the gold is slowly increased (Figure 1(b)). Eventually, the conductance drops as gold atoms migrate out of the constriction region[44]. If the rate of change of the conductance exceeds a pre-set target, the voltage across the gold bowtie is rapidly reduced to prevent catastrophic breaking of the gold junction, and then the voltage is slowly increased again. This sequence is repeated until the device conductance is reduced to a desired level at which point the voltage is reduced to zero.

An important advantage of the metallic etch mask is that the lateral size of the constriction can be estimated through current-voltage measurements, even for constrictions with dimensions below 10 nm, a scale where accurate AFM measurements become problematic because of non-zero AFM tip curvature. Specifically, the initial conductance in Fig. 1(b) is predominantly due to the resistance of the Au bowtie structure ($R_L$~435 Ω), and the decrease in conductance reflects narrowing of the constriction. While the exact conductance of the constriction is convolved with the conductance of the un-etched graphene, the final conductance value of ~1.8 mS can be used to construct an upper bound for the size of the etch mask. Assuming that electron scattering in the gold constriction is weak, each Au atomic channel (~0.3nm x 0.3nm) contributes one quantum of conductance, $G_0$ = ~1.8 mS, implying a maximal cross-section of ~9 $nm^2$ after FCE– a size beyond the reach of standard nanolithography.

Following FCE, an oxygen plasma etch is used to remove graphene regions that are unprotected by the gold. The plasma etch is relatively gentle, using a setting of 20 Watts power for 6 seconds in a capacitively coupled plasma etcher (Technics PEII-A). We find that gold that is narrowed to ~3nm width or less fails to act as a reliable etch mask and frequently results in disconnected graphene. After the plasma etch, the sample is immersed in a $KI/I_2$ gold etch solution for 30 seconds and subsequently cleaned with acetone and isopropanol. The result is a bowtie shaped structure of graphene with a nanometer scale constriction at the center. The large-area Pd/Cr contacts laid down in the second e-beam lithography step serve as source and drain electrodes, while the p++ doped silicon substrate is used as a back gate (Figure 1(a)).

Electrical measurements are performed using a Keithley 6517A and a NI-6221 DAQ card, which also drives the electromigration.

Graphene Raman spectra are obtained by exciting with a 514 nm wavelength laser under a 100× objective. Laser power is below 4 mW to avoid damage to graphene. The laser spot size is ~1μm in diameter. Single layer graphene can be identified by their unique Raman signatures, i.e. SLG has the Stokes G peak at 1583 $cm^{-1}$ and a single symmetric 2D band around 2700 $cm^{-1}$.

*Acknowledgment*

This work was supported by the National Science Foundation through grant #DMR-0805136 and partially by the Nano/Bio Interface Center (National Science Foundation

NSEC DMR-0425780) .  Use of facilities of the Nano/Bio Interface Center is gratefully acknowledged.

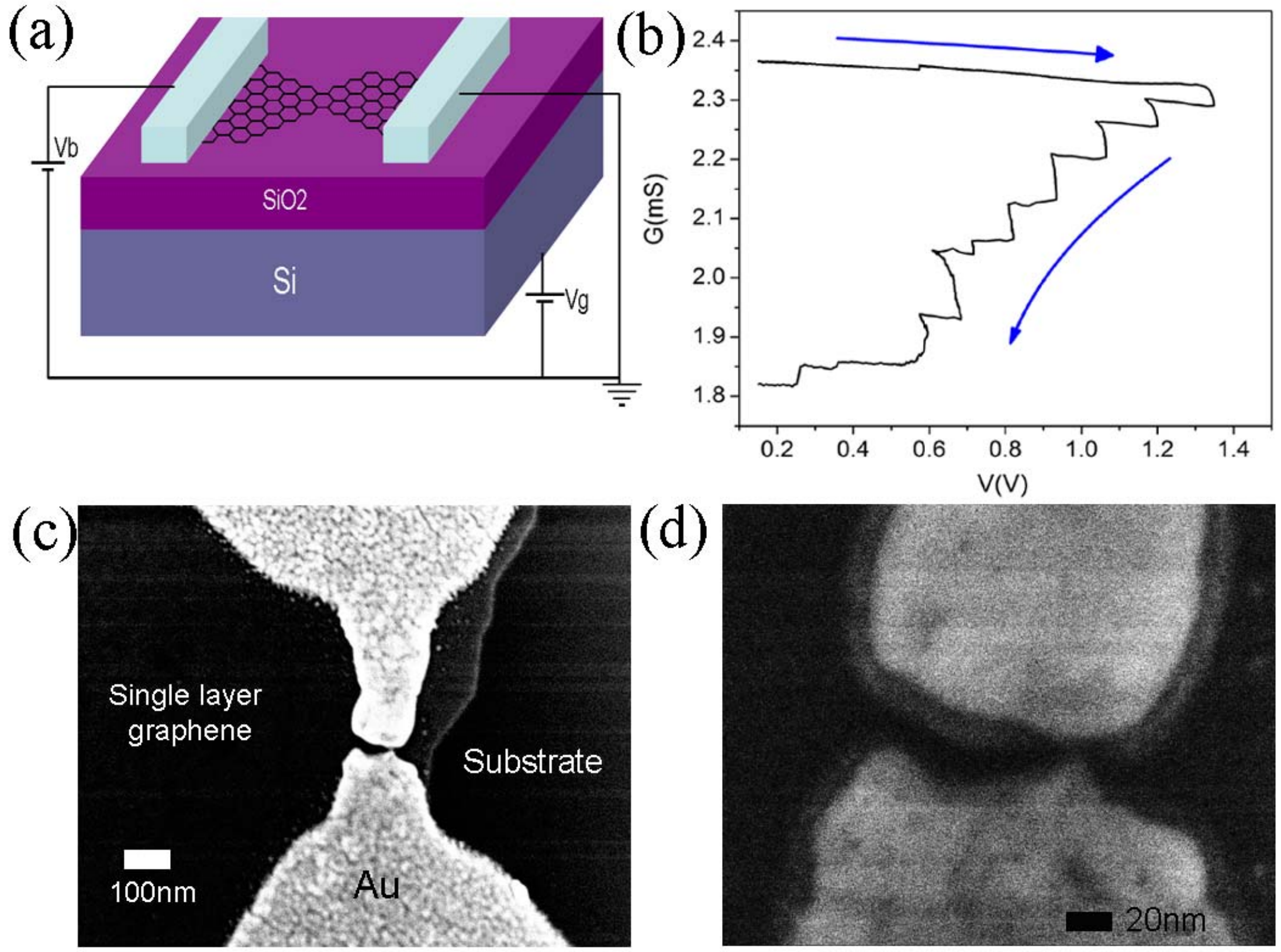

**Figure 1** (a) Diagram of the graphene nanoconstriction field effect transistor, constructed on a p++ doped Si wafer with 300 nm thermal oxide. The wafer serves as a back gate for the device, while the conductance of the graphene is monitored via Pd/Cr electrodes. (b) Conductance versus voltage G(V) curve obtained during FCE of a gold nanowire on graphene in vacuum. Blue arrow indicates how G(V) evolves as FCE proceeds (c) SEM image of a device after FCE. Bright regions in the image are gold. The left side of the image is covered with monolayer graphene, with an edge visible on the right. (d) High mangification image of the same device. The width of the gold nanowire on top of the graphene has been reduced to less than 10nm. The thinned gold is fainter in the image than the un-thinned electrode.

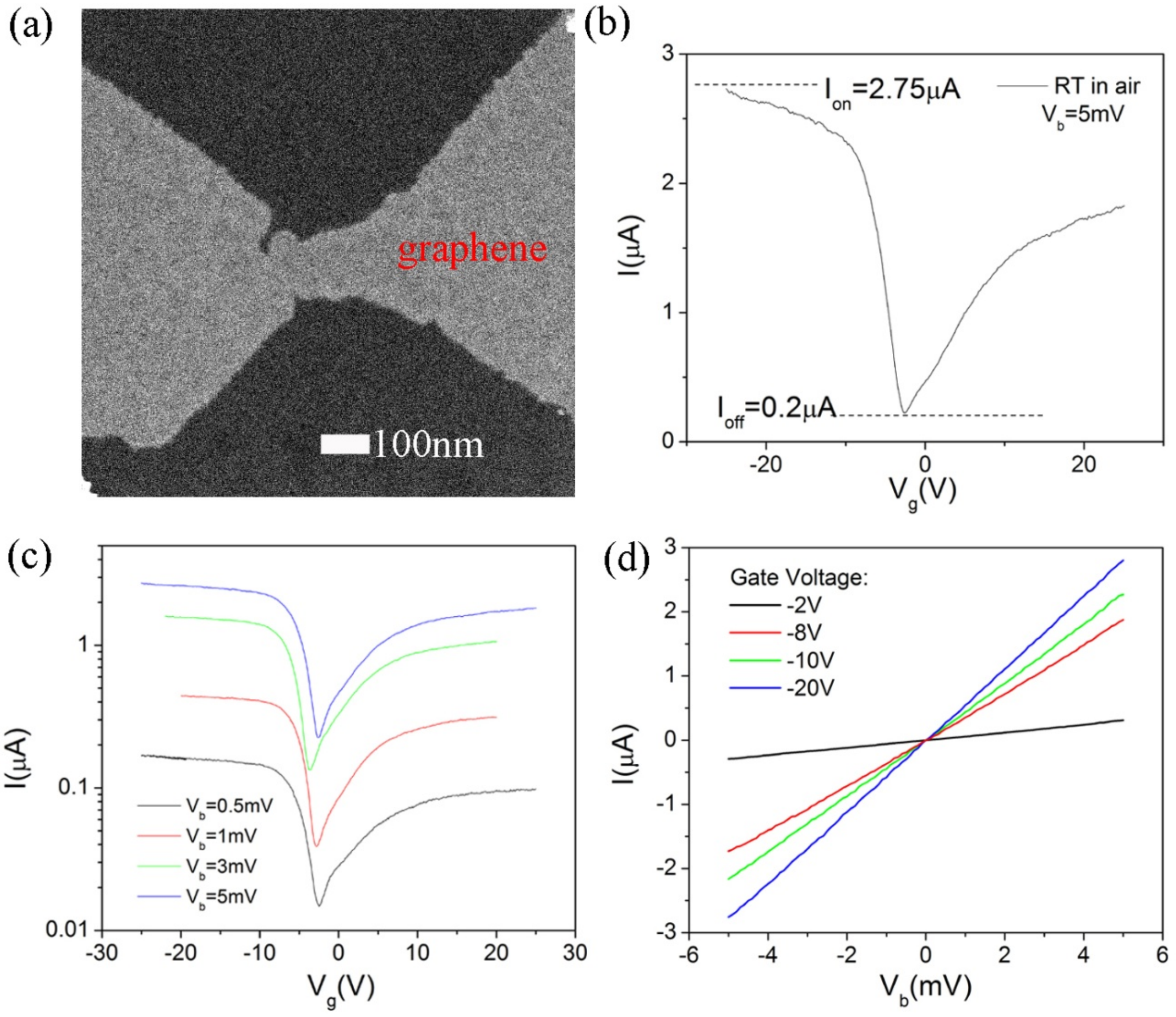

**Figure 2**. (a) SEM image of the a graphene NCFET device with a width of 90±5 nm. (b) Current versus gate voltage I (Vg) curve for the same device, which shows a current on/off ratio of 14 at room temperature. (c) I (Vg) characteristics of the same device at various bias voltages. (d) Current vs. bias characteristics for the same device at various gate voltages.

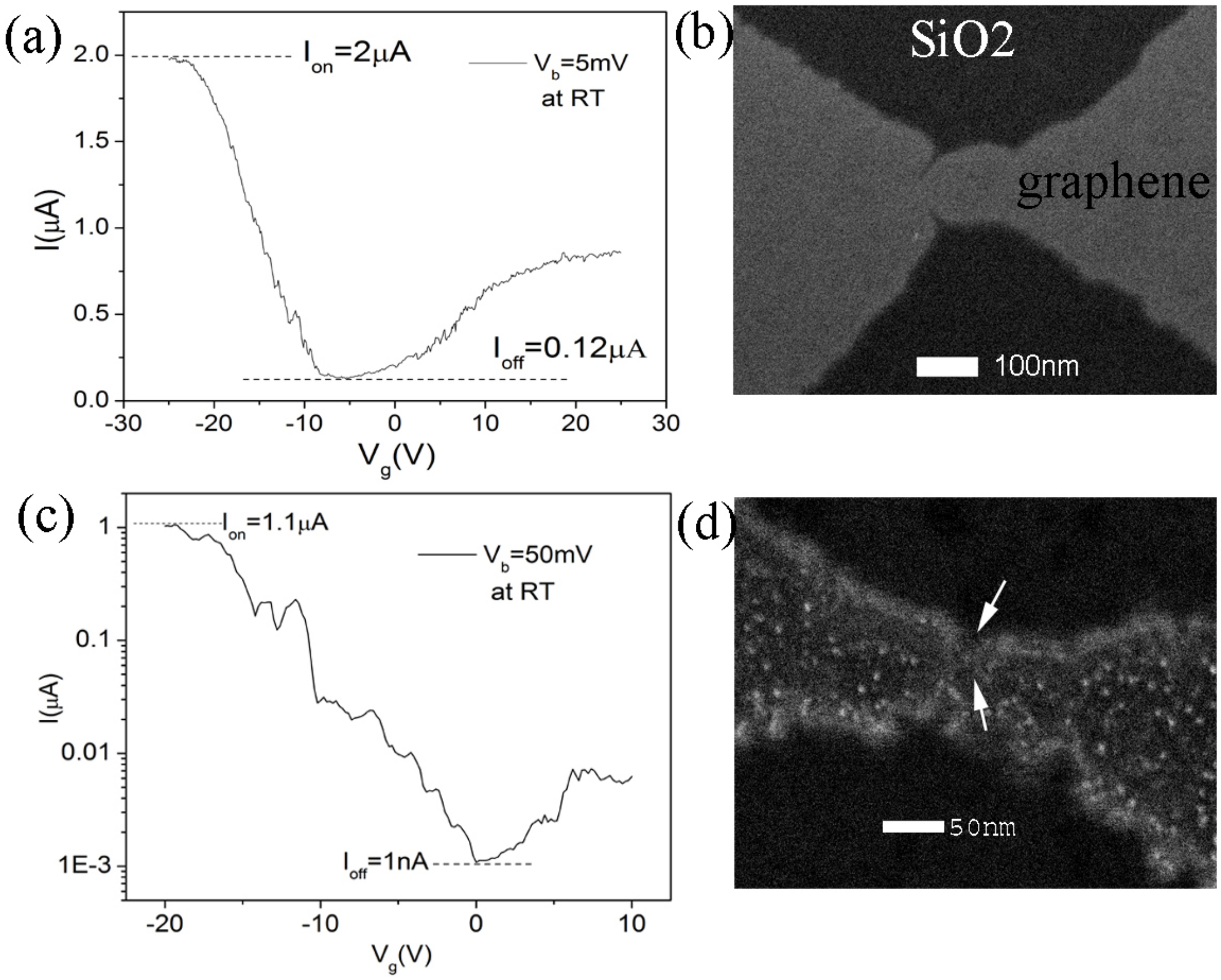

**Figure 3**. (a) Current versus gate voltage I(Vg) curve for a 37±3 nm wide NCFET device at room temperature. (b) SEM image of the device. (c) I (Vg) for a 9±3 nm wide NCFET device at room temperature. (d) SEM image with the constriction region indicated by white arrows.

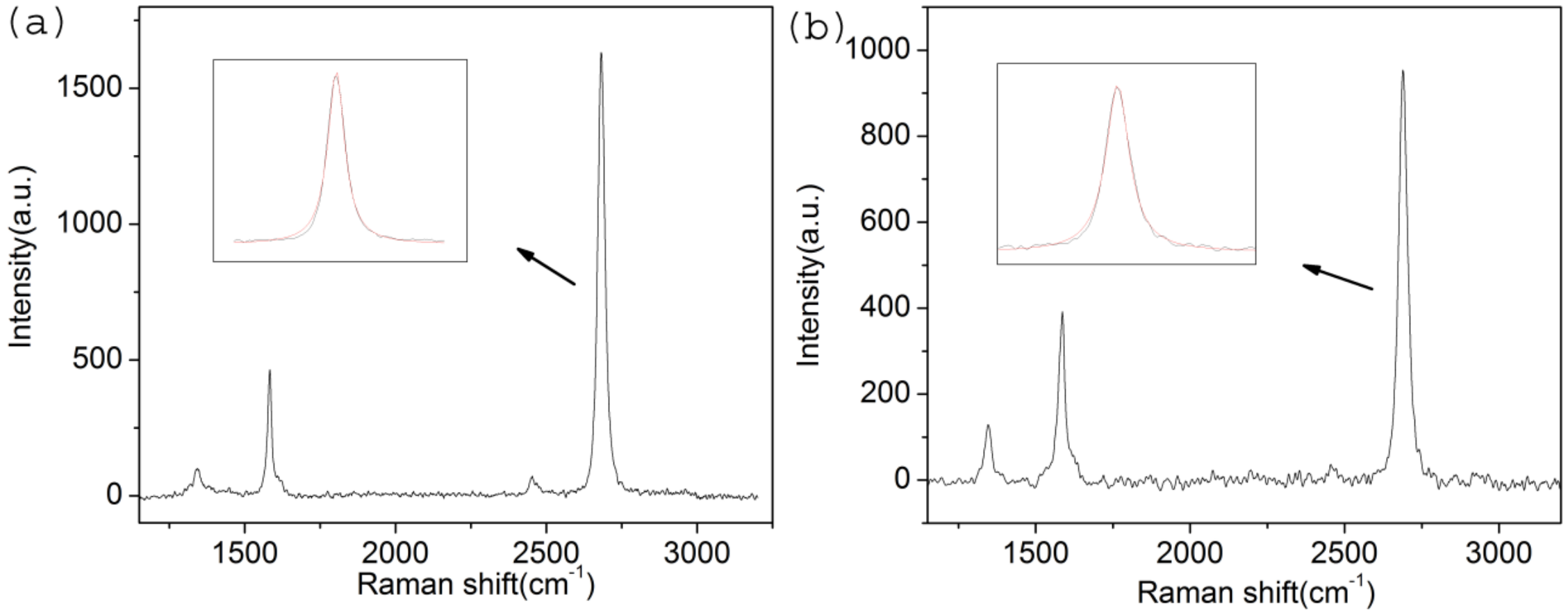

Figure 4. (a) Raman data and a Lorentz fit of the 2D peak (inset) for the devices shown in Figure 3(a/b); (b) Raman data and a Lorentz fit of the 2D peak (inset) for the devices shown in Figure 3(c/d) after all the other measurements. Both Raman show a large I(2D) (near 2700 $cm^{-1}$) /I(G) (near 1600 $cm^{-1}$) ratio, which is indicative of monolayer graphene.

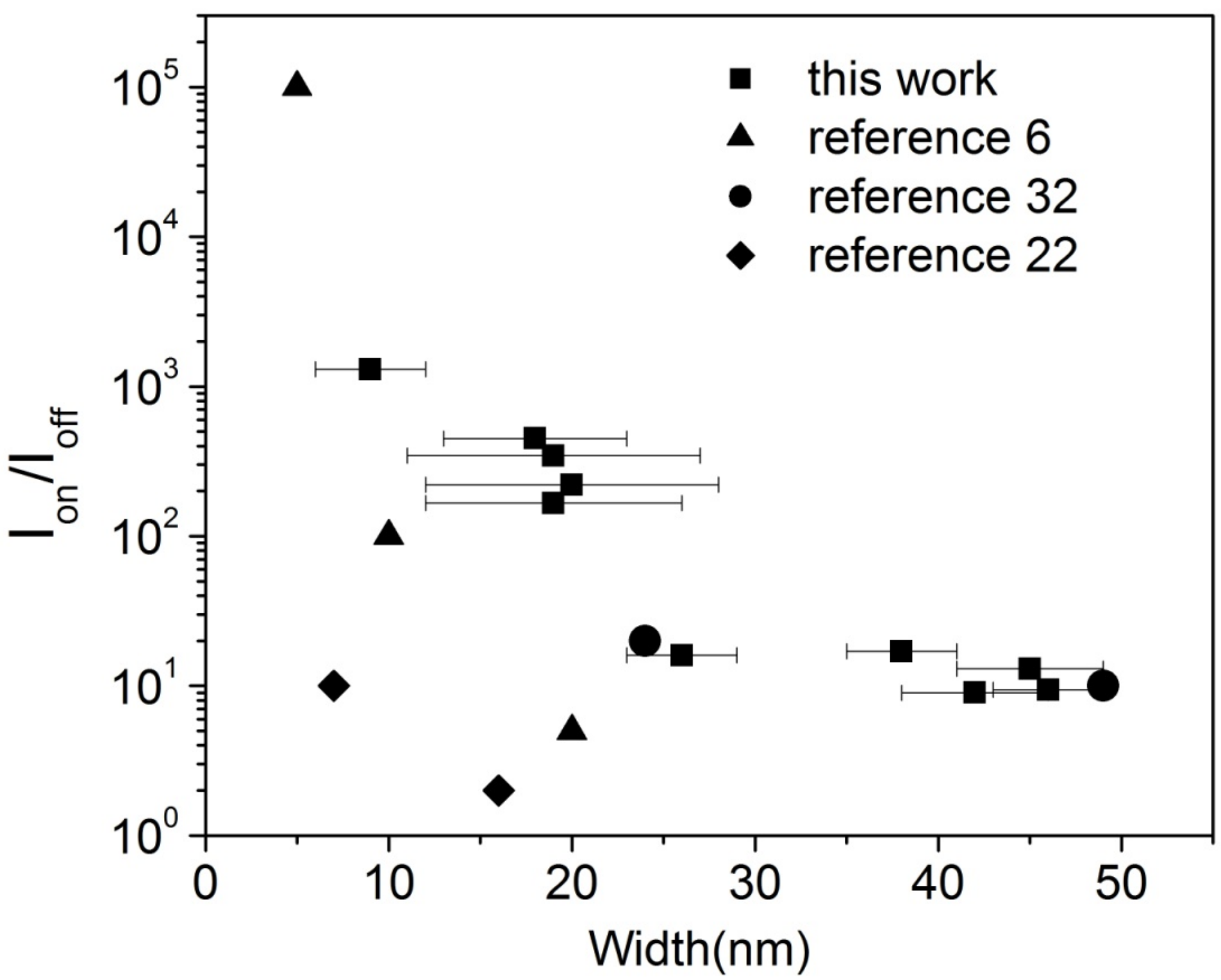

**Figure 5**. Room temperature current on/off ratios vs. constriction widths of graphene NCFET devices, in comparison with some other work.

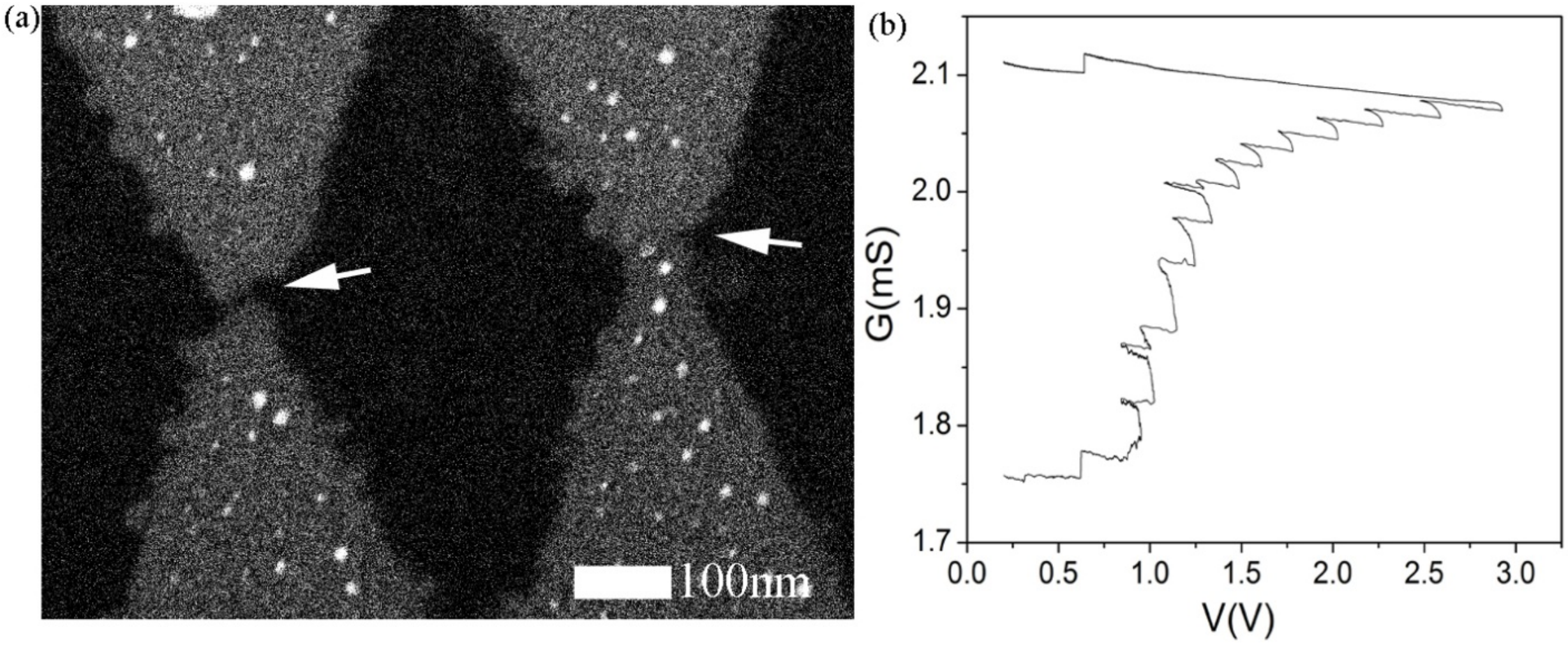

**Figure 6.** (a) SEM image of two NCFETs fabricated simultaneously on one piece of monolayer graphene. In the fabrication process, two Au tow-tie structures were fabricated on one piece of graphene, but only one single FCE process was performed, which thinned down Au structures simultaneously.  (b) The conductance vs. voltage characteristic of the FCE process of the same device.